\documentclass[useAMS,usenatbib,fleqn]{mn2e}

\usepackage{graphicx}
\usepackage{float}
\usepackage{amsmath}
\usepackage{amssymb}
\usepackage[labelformat=empty,labelsep=none]{subcaption}
\usepackage{epsfig}

\def\gtsim {\gtrsim}   
\def\ltsim {\lesssim}   

\newcommand{\ef}{\epsilon_\textrm{f}}
\newcommand{\ms}{M_\textrm{seed}}
\newcommand{\rc}{\rho_\textrm{c}}

\title[The Effects of AGN Feedback]{The Effects of AGN Feedback on Present-Day Galaxy Properties in Cosmological Simulations}
\author[P.~Taylor and C.~Kobayashi]{Philip~Taylor\thanks{E-mail:
p.taylor7@herts.ac.uk} and Chiaki~Kobayashi\\
Centre for Astrophysics Research, Science and Technology Research Institute, University of Hertfordshire, AL10 9AB, UK}

\begin{document}

\date{Accepted  Received ; in original form}

\pagerange{\pageref{firstpage}--\pageref{lastpage}} \pubyear{}

\maketitle

\label{firstpage}

\begin{abstract}
We show that feedback from active galactic nuclei (AGN) plays an essential role in reproducing the down-sizing phenomena, namely: the colour-magnitude relation; specific star formation rates; and the $\alpha$ enhancement of early type galaxies.
In our AGN model, black holes originate from Population {\sc iii} stars, in contrast to the merging scenario of previous works.
In this paper, we show how the properties of present-day galaxies in cosmological chemo-hydrodynamical simulations change when we include our model for AGN feedback.
Massive galaxies become redder, older, less massive, less compact, and show greater $\alpha$ enhancement than their counterparts without AGN.
Since we reproduce the black hole mass and galaxy mass relation, smaller galaxies do not host a supermassive black hole and their star formation history is affected very little, but they can get external enrichment from nearby AGN depending on their environment.
Nonetheless, the metallicity change is negligible, and the mass--metallicity relations, which are mainly generated by supernova feedback at the first star burst, are preserved.

\end{abstract}

\begin{keywords}
black hole physics -- galaxies: evolution -- galaxies: formation -- methods: numerical -- galaxies: abundances
\end{keywords}


\section{Introduction}
\label{sec:intro}

The $\Lambda$ Cold Dark Matter ($\Lambda$CDM) cosmology is the currently accepted best explanation for the structure we see in the Universe, and its evolution across cosmic time.
On large scales, structure formation is driven by the gravitational influence of dark matter, leading to Mpc-scale filaments, walls, and  voids by the present day.
Galaxies form within these, and their small-scale evolution is driven primarily by baryonic processes.
In the $\Lambda$CDM paradigm, all structure forms hierarchically; small halos collapse first, and merge over time to form ever larger ones.
This has been challenged observationally, however, by the down-sizing phenomenon \citep[e.g.,][]{cowie96,juneau05,bundy06,stringer09}, whereby the mass of the most rapidly star-forming galaxies decreases with time.
Similar phenomena have also been observed as the evolution in mass-to-light ratios \citep{treu05}, and the enhancement of $\alpha$ elements \citep{thomas03} in early-type galaxies.
In hydrodynamical simulations, many galaxies that form at high redshift merge to form the most massive galaxies today, and so these galaxies tend to have older populations of stars than present-day low-mass galaxies (\citet{ck07}; see also \cite{delucia06,fontanot09} for semi-analytic models).
With AGN feedback included, the star formation rate (SFR) at present can be reduced in line with observations \citep{pt14}, which will help reproduce the observations of the downsizing phenomenon.

The potential importance of super massive black holes (BHs) in galaxy evolution has been made clear by the discovery that black hole mass correlates with numerous properties of its host galaxy, including: bulge mass \citep{magorrian98,marconi03,haring04,sani11}; stellar velocity dispersion \citep{ferrarese00,gebhardt00,tremaine02,kormendyho13}; luminosity \citep{kormendy95,marconi03}; and S\'erisc index / stellar concentration \citep{graham01,graham07,savorgnan13}.
These correlations have been interpreted as implying that galaxies and BHs coevolve \citep[but see also][]{jahnke11}.
Such observed relations are reproduced by simulations \citep{croton06,dubois12,pt14}, as well as a better reproduction of the cosmic SFR \citep[e.g.,][]{booth09,pt14,vogelsberger14}.

We proposed a new model for AGN feedback in cosmological simulations \citep{pt14}.
With our AGN model, BHs form earlier in our simulations ($z\sim16$) and with a lower mass ($\sim10^3M_\odot$) than previous works ($\sim10^5M_\odot$).
This means that the effects of AGN feedback are different, especially at high redshift.
However, the peak in cosmic SFR density still occurs at $z=2$, consistent with observations and other theoretical works.
Indeed, the majority of stars in galaxies form before feedback from AGN becomes dominant, and so the scaling relations of galaxies are only weakly affected by AGN feedback.
We will illustrate this with the mass-metallicity relations (MZR), which is the most stringent constraint on galaxy evolution.
The galaxy size--mass relation is more affected, with AGN feedback increasing galaxy effective radius, but is still consistent with observations (but see also \citet{dubois13} who find that AGN feedback is necessary to match observations).

In this paper, we examine the effects of our AGN feedback model on the chemodynamical evolution of galaxies in a cosmological context.
We compare the $z=0$ properties of galaxies in two cosmological simulations, one with, and one without, AGN with observations, and  look in particular at down-sizing and galaxy scaling relations.
In Section \ref{sec:model}, we introduce the code used to perform our simulations, and briefly describe our prescriptions for stellar and AGN feedback.
We outline the simulations performed in Section \ref{sec:sims}, as well as describing in more detail three example galaxies that will be used to highlight the differences in the simulations.
Our results are presented and discussed in Section \ref{sec:res}, and we draw our conclusions in Section \ref{sec:conc}.


\section{Model}
\label{sec:model}

We use the smoothed particle hydrodynamics (SPH) code {\small GADGET-3} \citep{springel01gadget,springel05gadget}, which is fully adaptive with individual smoothing lengths and timesteps, and uses an entropy-conserving formulation of SPH \citep{springel02}.
This is augmented with various physical processes relevant to the formation and evolution of galaxies:  metal-dependent radiative cooling, star formation, stellar feedback \citep{ck07} from AGB stars, Type Ia, and collapse supernovae and hypernovae, and BH physics \citep{pt14}.
The BH physics is described in detail in \citet{pt14}; we summarize the key points below.

\subsection{Black Hole Physics}
\label{sec:modelbh}

i) BHs are {\bf seeded} with a mass $\ms$ from gas particles of primordial composition (i.e. $Z=0$) that are denser than some defined critical density (i.e. $\rho_\textrm{g}>\rc$).
In \citet{pt14}, we found that values of $\ms=10^{2-3}\,h^{-1}M_\odot$ and $\rc=0.1\,h^2m_\textrm{H}\,\textrm{cm}^{-3}$ were needed in order to produce results in agreement with observational constraints (specifically, cosmic SFR density, the $M_{\rm BH}$--$\sigma$ relation, and the galaxy size--mass relation).
Such a seed mass is consistent with the idea that the supermassive black holes observed today are the progeny of Population {\sc iii} star remnants, in keeping with the lack of observational evidence of chemical enrichment from pair-instability supernovae \citep[e.g.,][]{ck11dla}.

Due to the limited mass resolution, $\ms$ is much smaller than the mass of gas particles, $M_\textrm{gas}$.
For this reason, and in keeping with previous works \citep[e.g.,][]{springel05,booth09}, we store the black hole mass separately, and use $M_\textrm{gas}$ as the dynamical mass until $M_\textrm{BH}>M_\textrm{gas}$.
Additionally, due to the spatial resolution, it is possible for BHs to migrate out of galaxies in a few timesteps due to numerical effects.
To avoid this, we calculate the centre of mass of all particles within the smoothing length of a BH in each timestep, and reposition the BH to that location.

ii) BHs {\bf grow} via both gas accretion and mergers.
Given the spatial resolution of our simulations, it is necessary to adopt a sub-resolution model for both the growth of, and feedback from, BHs.
To estimate the gas accretion rate, we use the Bondi-Hoyle formula \citep{bondi44} multiplied by a constant factor $\alpha$ that accounts for the finite resolution of the simulations.
In \citet{pt14}, we treated $\alpha$ as a free parameter, and chose a value of 1 from observational constraints.
$\dot M_{\rm acc}$ could be estimated by more sophisticated methods \citep[e.g.,][]{power11,angles13}, but higher resolution simulations are required than in this work.
We assume that accretion is Eddington-limited at all times, and that BHs have a radiative efficiency of 0.1 \citep{shakura73}.

Black holes also grow via mergers with one another.
A pair of BHs merge if their relative speed is less than the local sound speed, and their separation is smaller than the gravitational softening length of the black holes.
We use the sound speed as a measure of the local velocity scale, which precludes black holes that are undergoing a quick flyby from merging \citep{springel05,dimatteo08}.

iii) The {\bf feedback} energy that a BH produces is proportional to the accretion rate, of which a constant fraction $\ef$ couples to neighbouring gas particles.
With our AGN model, we found that $\ef=0.25$ is required to reproduce observations \citep{pt14}, while previous works have used $0.05<\ef<0.15$ \citep[e.g.,][]{springel05,dimatteo08,booth09}.
The energy is distributed isotropically, kernel-weighted, and in a purely thermal form to a constant number of neighbouring gas particles, $N_\textrm{FB}=72$ (see \citet{pt14} for more discussion).

\subsection{Stellar Physics}
\label{sec:modelstars}

Photo-heating is given by a uniform and time-evolving UV background radiation \citep{haardt96}, and radiative cooling is computed using a metallicity-dependent cooling function \citep{sutherland93}.
The star formation criteria used are the same as in \citet{katz92}: (1) converging flow, $(\nabla \cdot \mbox{\boldmath$v$})_i<0$; (2) rapid cooling, $t_\textrm{cool}<t_\textrm{dyn}$; and (3) Jeans unstable gas, $t_\textrm{dyn}<t_\textrm{sound}$.
The star formation timescale is taken to be proportional to the dynamical timescale ($t_{\rm sf} \equiv \frac{1}{c_*}t_{\rm dyn}$), where $c_*$ is a star formation timescale parameter, which we set to $0.02$.
We spawn new star particles as in \citet{ck07}, and follow the evolution of the star particles at every timestep.
The energies are distributed to the same number of neighbouring gas particles as for AGN feedback, weighted by an SPH kernel.
The energies of mass loss and supernovae are distributed in purely thermal form, although a fraction of it could, in principle, be distributed in kinetic form as a velocity perturbation to the gas particles \citep{navarro93}.

The chemical enrichment is computed with the scheme of \citet{ck04}, which does not include the instantaneous recycling approximation.
Different from \citet{ck07}'s simulations, we adopt the initial mass function (IMF) of stars from \citet{kroupa08}, and the updated neucleosynthesis yields of \citet{ck11b} for $1-50 M_\odot$ stars.
Note that the mass loss from low-mass stars has been included since \citet{ck04}.
The effects of hypernovae are included with a metal-dependent hypernova fraction as in \citet{ck11a}.
The progenitor model of Type Ia supernovae (SNe Ia) is very important for predicting [$\alpha$/Fe] of galaxies because SNe Ia produce more iron than $\alpha$ elements (O, Mg, Si, S, and Ca).
Our SN Ia model is based on the single degenerate scenario with the metallicity effects of white dwarf winds \citep{ck98,ck00}, and the lifetime distribution functions are calculated as in \citet{ck09}.
These provide an excellent agreement with the observed elemental abundances in the solar neighbourhood \citep{ck11a}.
As a result, half of iron is produced by SNe Ia, and the rest by core-collapse supernovae.

\section{The Simulations}
\label{sec:sims}

In this section, we describe the set-up and large-scale evolution of our simulations, before then describing the properties of three galaxies that will be used for illustrative purposes in Section \ref{sec:res}.

\subsection{Cosmic Evolution}
\label{sec:cosbox}

\begin{figure*}
\centering
\includegraphics[width=\textwidth,keepaspectratio]{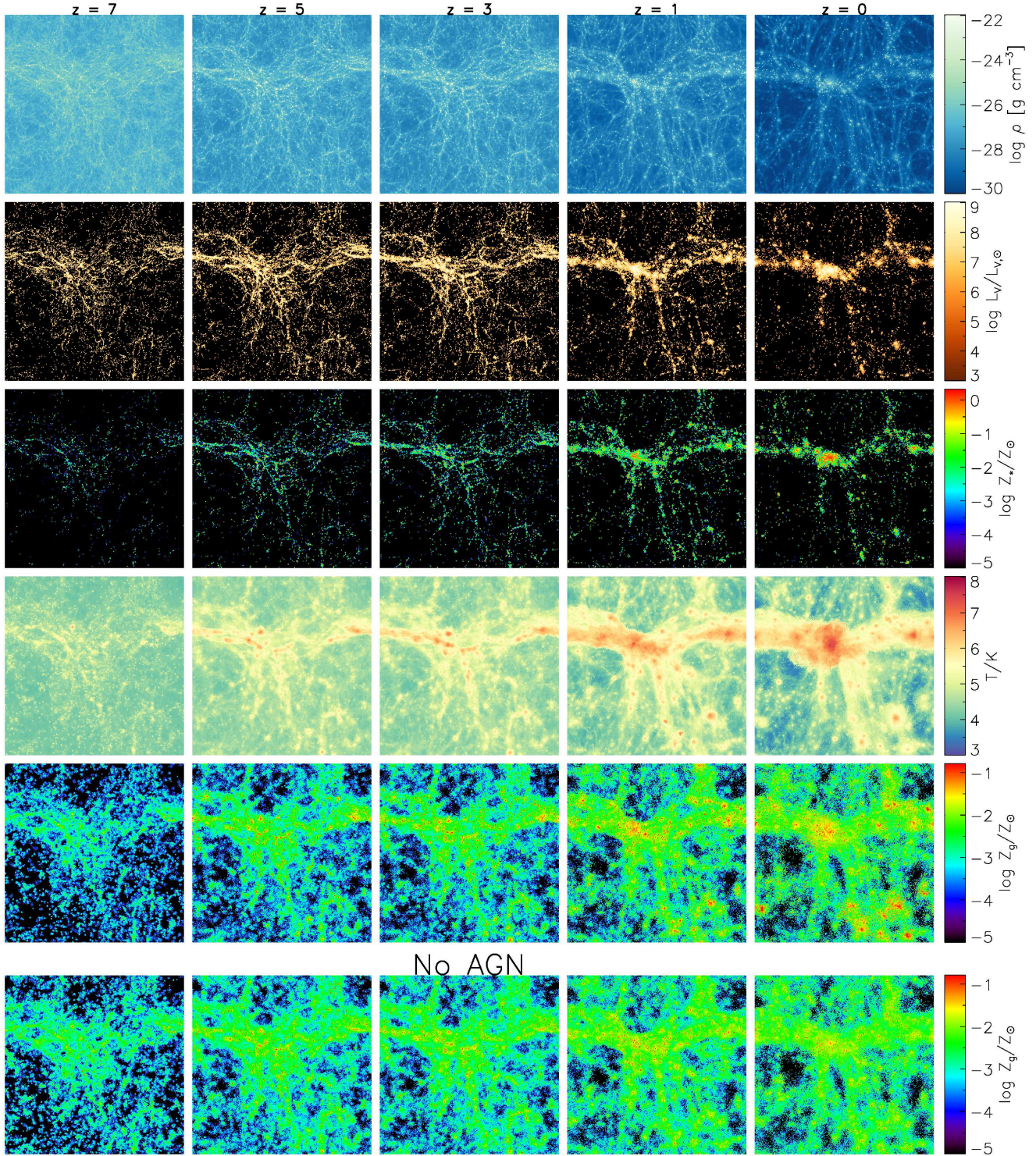}
\caption{Evolution of cosmological simulation in  $25h^{-1}$Mpc box with (top 5 rows) and without (final row) AGN feedback included.
We show projected gas density on the upper row, rest frame $V$ band luminosity and stellar metallicity on the second and third, gas temperature on the fourth, and gas metallicity on the final two rows.}
\label{fig:simevol}
\end{figure*}


We present simulations run both with and without the AGN model described in Section \ref{sec:modelbh}.
With AGN feedback included, the fiducial model parameters of \citet{pt14} are used, namely $\alpha=1$, $\ef=0.25$, $\ms=10^3h^{-1}M_\odot$, and $\rc=0.1h^2m_\textrm{H}{\rm cm}^{-3}$.
The simulations in this paper are run at the same resolution with identical cosmological initial conditions.
We employ a WMAP-9 $\Lambda$CDM cosmology \citep{wmap9} with $h=0.70$, $\Omega_\textrm{m}=0.28$, $\Omega_\Lambda=0.72$, $\Omega_\textrm{b}=0.046$, and $\sigma_8=0.82$.
The initial conditions are in a periodic, comoving, cubic volume 25 $h^{-1}$ Mpc on a side, are chosen to have a central cluster at $z=0$, and are different from \citet{ck07}, which corresponds to a typical field of the universe.
They contain $240^3$ particles of each of dark matter and gas, with masses $M_\textrm{DM}=7.3\times 10^7M_\odot$ and $M_\textrm{gas}=1.4\times 10^7M_\odot$.
By $z=0$ there are approximately 3 per cent more particles in total when AGN feedback is not included due to the formation of star particles, and the influence of AGN feedback.
We use gravitational softening lengths of $\epsilon_{\rm DM}=2.25h^{-1}$ kpc and $\epsilon_{\rm gas}=1.125h^{-1}$ kpc for dark matter and gas respectively.
The finite volume of the simulation means that we do not necessarily obtain a representative sample of the most massive galaxies, while the low-mass end may be affected by the limited resolution of the simulations.
When comparing to observations in the subsequent sections, our simulated galaxies are complete for $9\ltsim\log M/M_\odot \ltsim 11.4$, $-25.6\ltsim M_{\rm K}\ltsim -21$, and $-22.1\ltsim M_{\rm B}\ltsim 4(B-V)_0-19$, although we compare our simulations to each other across a wider range since their initial conditions are the same.

The redshift evolution of the simulation with AGN is shown in Fig. \ref{fig:simevol}, as well as the gas metallicity in the simulation without AGN.
In both cases, a rich filamentary structure exists (top row) at high redshift, to which star (second row) and black hole formation is mostly confined.
Without AGN, the first star forms at $z\sim14$ in the region that will subsequently collapse to form the largest cluster in the simulation box.
As early as $z=7$, supernova feedback heats (fourth row) and chemically enriches the gas.
This enrichment enhances star formation by allowing the gas to cool more rapidly.
The cosmic SFR then peaks at $z\sim2-3$.
Supernova feedback causes galactic winds to be produced, which cause the intergalactic medium (IGM) to become enriched by interstellar medium (ISM; sixth row).

With AGN feedback included, BHs start forming from $z>20$, and prevent stars from forming until $z\sim11.5$.
When these first stars do form, however, they enrich the surrounding gas and inhibit further BH formation.
As in the case without AGN, the cosmic SFR peaks at $z\sim 2$, coincident with a broader peak in the BH accretion rate (Fig. 2 of \citet{pt14}).
The BH number density peaks earlier, at $z\sim5.5$, after which mergers reduce their number.
AGN and supernova feedback together produce stronger galactic winds than supernova feedback alone, and a larger region of the IGM is influenced (fifth row).
The largest concentration of stars and BHs is seen in the central cluster of galaxies, with a present total mass of $\sim10^{14}h^{-1}M_\odot$, and which hosts the most massive black hole in the simulation with a mass of $\sim2\times10^8h^{-1}M_\odot$.

\begin{figure}
\centering
\includegraphics[width=0.47\textwidth,keepaspectratio]{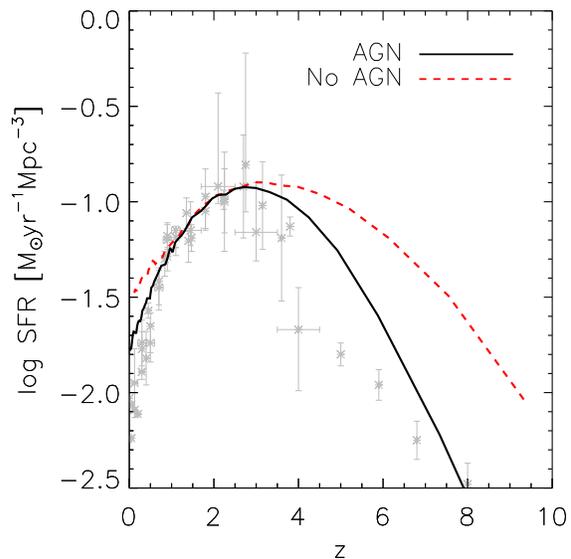}
\caption{Cosmic star formation rate density as a function of redshift for the simulations with (solid black line) and without (dashed red line) AGN feedback.
Sources of observational data (grey points) are listed in Section \ref{sec:cosbox}.}
\label{fig:cosmicsfr}
\end{figure}

We show in Fig. \ref{fig:cosmicsfr} the cosmic SFR density for the simulations with (solid black line) and without (dashed red line) AGN feedback.
Observational data are plotted in grey, and are taken from \citet{bouwens11}, \citet{karim11}, \citet{cucciati12}, \citet{oesch12}, \citet{burgarella13}, \citet{gunawardhana13}, and \citet{sobral13}, adjusted to the Kroupa IMF \citep{bernardi10}; see \citet{pt14} for a discussion on the dust extinction and IMF dependence.
Both simulations show a peak in SFR of similar magnitude at $z\sim2-3$, in agreement with observations.
At higher and lower redshifts, however, the inclusion of AGN significantly reduces the SFR, and the simulated data lie much closer to the observations.
At high redshift, although the BHs have not yet grown to be supermassive, gas has not been very much enriched, and so it is unable to cool efficiently.
By low redshift, the supermassive BHs are able to accrete a large amount of gas, and provide enough energy to surrounding gas to drive powerful galactic winds.

\subsection{Individual Galaxies}
\label{sec:abc}

\begin{figure*}
\centering
\includegraphics[width=\textwidth,keepaspectratio]{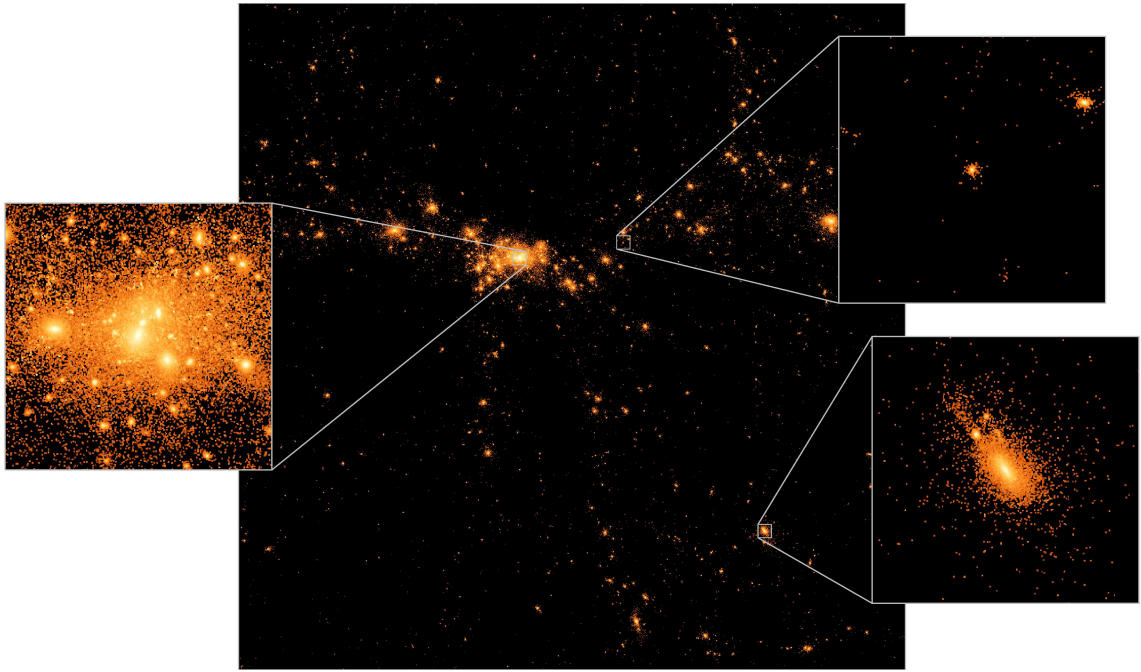}
\caption{Integrated V band luminosity image of the simulation with AGN feedback.
The zoom-ins are $400h^{-1}$kpc on a side, and show the locations of galaxies A (left), B (bottom right), and C (top right), as well as their immediate environment in the simulation box ($25h^{-1}{\rm Mpc}$).}
\label{fig:abcvband}
\end{figure*}
\begin{figure}
\centering
\includegraphics[width=0.48\textwidth,keepaspectratio]{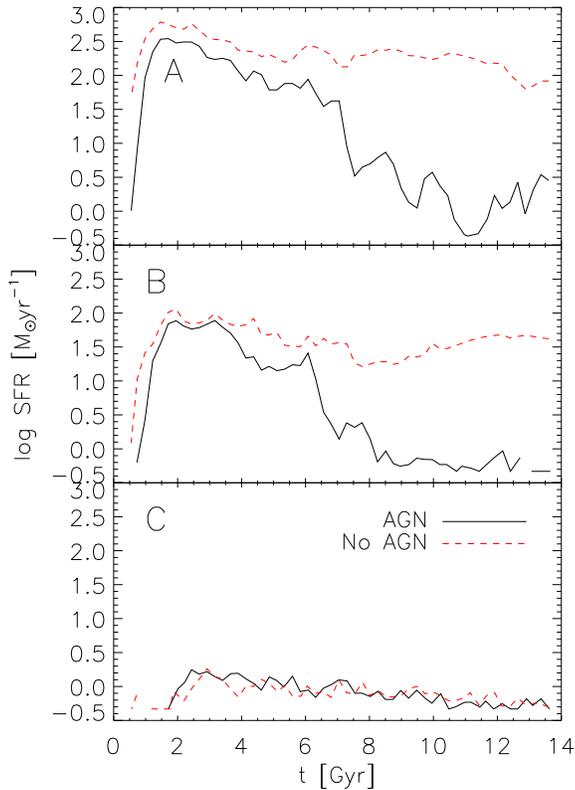}
\caption{Star formation rate as a function of time since the Big Bang for the three example galaxies A, B, and C (Section \ref{sec:abc}) in the simulations with (solid black lines), and without (dashed red lines) AGN feedback.}
\label{fig:sfrabc}
\end{figure}

\begin{table*}
\caption{$z=0$ properties of galaxies A, B, and C both with and without AGN feedback.
(1) Estimated from core-S\'ersic fit.
(2) Gas fraction within $R_{\rm e}$, calculated as $f_{\rm g}=M_{\rm g}/(M_{\rm g}+M_*+M_{\rm BH})$.
(3) $V$ band luminosity weighted within $R_{\rm e}$.
(4) SFR weighted within $R_{\rm e}$.
Note that there is no star-forming gas in A when AGN is included.
(5) Galaxy colour in central $1h^{-1}{\rm kpc}$.}
\begin{tabular}[width=\textwidth]{lcccccccccc}
	Galaxy & $M_*$ & $M_{200}$ & $R_{200}$ & $f_\textrm{g}$ & $<$Age$>$ & [M/H]$_*$ & [O/H]$_\textrm{g}$ & [O/Fe]$_*$ & $(B-V)_0$ &  $\log(M_\textrm{BH})$\\
	& $h^{-1} {\rm M}_\odot$ & $h^{-1} {\rm M}_\odot$ & $h^{-1}$ kpc & & Gyr & & & & & $h^{-1} {\rm M}_\odot$\\
	& (1) & & & (2) & (3) & (3) & (4) & (3) & (5) & \\
	\hline
	A (AGN) & 6.8e11 & 9.5e13 & 743.2 & 0.001 & 9.71 & 0.23 &   -   & 0.18 & 0.92 & 8.4\\
	A (No AGN) & 1.7e12 & 9.9e13 & 752.6 & 0.03 & 2.19 & 0.35 & 0.39 & -0.10 & 0.41 & -\\
	B (AGN) & 1.4e11 & 5.1e12 & 279.1 & 0.01 & 10.09 & 0.10 & 0.03 & 0.20 & 0.92 & 7.9\\
	B (No AGN) & 3.7e11 & 6.0e12 & 295.8 & 0.05 & 2.90 & 0.33 & 0.41 & -0.10 & 0.37 & -\\
	C (AGN) & 5.4e9 & 1.0e11 & 76.5 & 0.17 & 4.79 & 0.15 & 0.49 & 0.05 & 0.72 & 5.3\\
	C (No AGN) & 4.8e9 & 1.0e11 & 77.4 & 0.22 & 3.08 & 0.28 & 0.35 & 0.01 & 0.54 & -\\
\end{tabular}
\label{tab:abc}
\end{table*}

Galaxies are identified using a Friend of Friends (FoF) algorithm \citep{springel01} as in \citet{pt14}, which produces a catalogue of groups of particles whose mass exceeds $32M_\textrm{DM}$.
For the analysis of properties of a galaxy, only those groups containing at least 50 star particles are used.
In the simulation with (without) AGN, 1380 (1009) groups are found by the Friends of Friends algorithm, of which 1125 (742) are classified as galaxies, corresponding to $10^9\lesssim M_*/M_\odot\lesssim10^{12}$.
These differences are caused by AGN feedback at high redshift; although weak, early feedback from BHs suppresses star formation  and low-mass galaxies survive.
With AGN feedback included, 957 galaxies contain at least 1 BH.

Galaxy stellar masses and effective radii are estimated by fitting a core-S\'ersic profile \citep{graham03,trujillo04} to the surface density.
Approximately 80 per cent of galaxies are well fit by such a function, and a further 10 per cent are well fit by a standard S\'ersic function.
In both cases we restrict the S\'ersic index to $n=4$; this gives a much better fit to our simulated galaxies than with a S\'ersic function and varying $n$ as in observations.
Galaxies that are not fit by either function tend to be near our resolution limit ($\sim10^9M_\odot$) or merging systems, and are excluded from our analysis.
Galaxy luminosities are estimated by applying the simple stellar population models of \citet{bc03} to star particles with a Chabrier IMF, which gives very similar results to the Kroupa IMF used in our simulations.

In Section \ref{sec:res} we will highlight the effects of AGN feedback by looking more closely at 3 example galaxies, matched between the runs with and without AGN.
These galaxies, which we refer to as A, B, and C, have stellar masses $\sim10^{12}$, $\sim10^{11}$, $\sim10^{9}M_\odot$ respectively, evolve in different environments, and are influenced differently by their BHs.
We show in Fig. \ref{fig:abcvband} images of the three galaxies at $z=0$ from the simulation with AGN feedback, as well as their locations in the cosmological box of our simulation, colour-coded by integrated $V$ band luminosity.
The global properties of these galaxies are summarized in Table \ref{tab:abc}.

Galaxy A is the most massive galaxy in the simulations, and sits at the centre of the cluster.
It experiences a major merger at $z\sim2$, and numerous minor mergers throughout its life.
With AGN, it hosts the most massive black hole, and AGN-driven outflows are produced intermittently.
AGN feedback heavily suppresses star formation in the centre of A, as shown in the top panel of Fig. \ref{fig:sfrabc}, although a small amount of star formation is ongoing in the outskirts of the galaxy ($\gtrsim20h^{-1}$kpc) at $z=0$ due to minor mergers.
As a result,  the mass- and luminosity-weighted stellar ages are older, and the stellar mass of A is reduced by a factor of about 2.5.
The removal of gas by AGN-driven winds reduces both $M_{200}$ and $R_{200}$, as well as the present gas fraction of this galaxy.
This also changes the distribution of heavy elements in this galaxy; stellar metallicity is slightly lower, though this difference is small enough that the mass-metallicity relations are not affected (Section \ref{sec:massmet}), and stars show $\alpha$-enhancement.

Galaxy B is a large field galaxy away from the main cluster (see Fig. \ref{fig:abcvband}).
It undergoes no major mergers, and very few minor ones.
With AGN, an outflow is produced at $z\sim2$, and by the present, star formation is fully quenched over the whole galaxy by AGN feedback (Fig. \ref{fig:sfrabc}) since, unlike in galaxy A, there is no regular gas supply.
There is an enhancement in both gas metallicity and temperature in a wider region of the IGM around B when AGN feedback is included.
Its stellar mass is diminished by just over half by $z=0$, and its gas mass is reduced from $1.5\times10^{11}M_\odot$ to $10^{10}M_\odot$, causing its gas fraction to be reduced from 0.13 to 0.03.
Both stellar and gas-phase metallicities are lower with AGN feedback, and stars are $\alpha$-enhanced.

Galaxy C lies on a filament near the main cluster (see Fig. \ref{fig:abcvband}) and is the smallest of the 3 galaxies.
It experiences a single merger at high redshift, but otherwise evolves passively.
AGN feedback has almost no effect on it; from the bottom panel of Fig. \ref{fig:sfrabc}, the star formation rate within this galaxy is similar for the two runs  across cosmic time, and the differences are negligible.
However, the chemical enrichment of its gas is enhanced with AGN feedback due to external enrichment from the strong, enriched winds generated in more massive galaxies.


\section{Results}
\label{sec:res}

In this section, we present the scaling relations of simulated galaxies in our cosmological simulations, one with the AGN feedback model of Section \ref{sec:modelbh} included, and one without.
Where applicable, we highlight the effects of AGN feedback by showing how the properties of galaxies A, B, and C (introduced in Section \ref{sec:abc}) change.

\subsection{Luminosity and Stellar Mass Functions}

\begin{figure}
\centering
\includegraphics[width=0.48\textwidth,keepaspectratio]{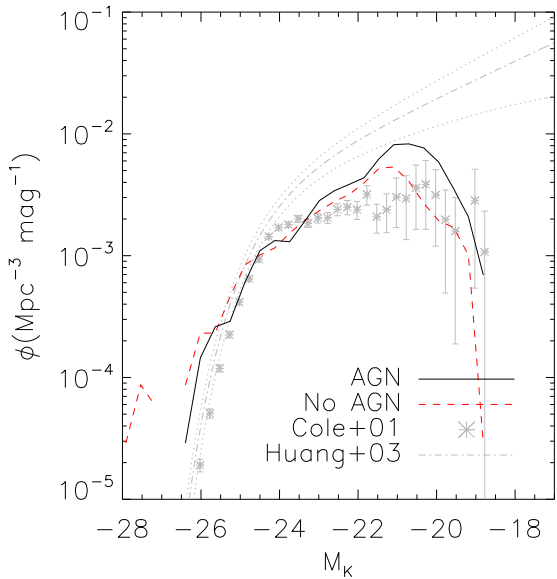}
\caption{Luminosity function for the simulations with (solid black line) and without (dashed red line) AGN feedback included.
Our calculated luminosity functions are compared with the observational data of \citet[][grey points]{cole01} and the Schechter function fit to the data of \citet[][grey dot-dashed line]{huang03}.}
\label{fig:lf}
\end{figure}

\begin{figure}
\centering
\includegraphics[width=0.48\textwidth,keepaspectratio]{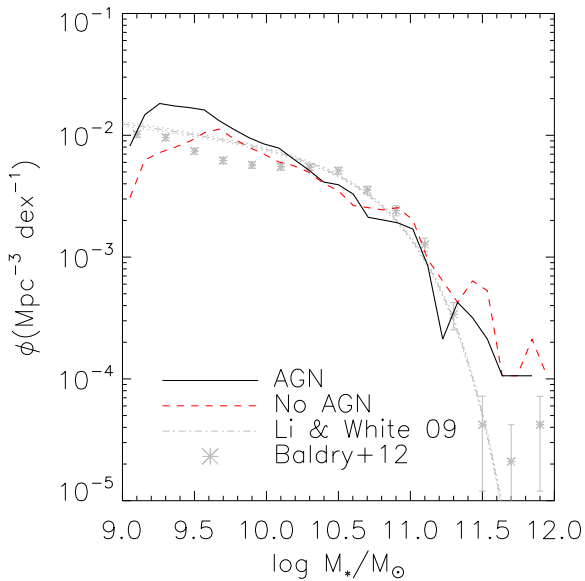}
\caption{Stellar mass function for the simulations with (solid black line) and without (dashed red line) AGN feedback.
Observational data are from \citet[][grey dot-dashed line]{li09} who use the GAMA survey, and \citet[][grey points]{baldry12} who use SDSS.}
\label{fig:mf}
\end{figure}


We show in Fig. \ref{fig:lf} the $K$ band galaxy luminosity functions for our simulations.
In the top panel, our calculated luminosity functions, both with (solid black line) and without (dashed red line) AGN feedback, are compared to the observational data of \citet[][grey points]{cole01} and the Schechter function fit to the data of \citet[][grey dot-dashed line]{huang03}.
For the completeness limit of our simulation $-25.6< M_{\rm K} <-20$, discussed in Section \ref{sec:cosbox}, both simulations show fairly good agreement with observations, though there are more low luminosity galaxies with AGN feedback (Section \ref{sec:abc}), and fewer high luminosity galaxies, compared to the simulation without AGN.
Looking more closely, we see that around the observed value of the Schechter function parameter $M_{\rm K}^*\sim-24.3$, both simulations produce very similar results, but are slightly lower than observed.
This may suggest that supernova, and possibly also AGN, feedback is slightly too strong around this mass  in our simulations.
The number of the most luminous galaxies ($M_{\rm K}\ltsim-25$) is most affected by AGN feedback, and is decreased by AGN feedback, but is still slightly higher than in observations.
This suggests that a larger simulation box, and possibly stronger AGN feedback at this mass, is required (see Section \ref{sec:conc} for more discussion).
Note that the agreement is not as good as in semi-analytic models \citep[e.g.,][]{bower06,guo13,schaye14} where the luminosity function is used to calibrate the parameters of the subgrid physics.  In hydrodynamical simulations, baryon physics is modeled on a particle basis with a smaller number of parameters.


Fig. \ref{fig:mf} shows the galaxy stellar mass functions for our simulations.
Both simulations show good agreement with the data of \citet{li09} and \citet{baldry12} for the completeness limit of our simulations $9\ltsim\log M/M_\odot\ltsim11.4$ (Section \ref{sec:cosbox}).
The difference between the simulations with and without AGN is not as pronounced at the high-mass end as for the luminosity function, but there are only a small number of galaxies with $\log M/M_\odot>11.5$.
The excess of low-mass galaxies when AGN feedback is included is due to the moderate suppression of SF early in the galaxies' lives (see Section \ref{sec:abc}).
Despite this, our mass functions show a better fit to the observational data than obtained from other hydrodynamical simulations \citep[e.g.,][see also Fig. 5 of \citet{schaye14}]{oppenheimer10,puchwein13,khandai14}.

\subsection{Colour -- Magnitude Relation}

\begin{figure}
\centering
\includegraphics[width=0.48\textwidth,keepaspectratio]{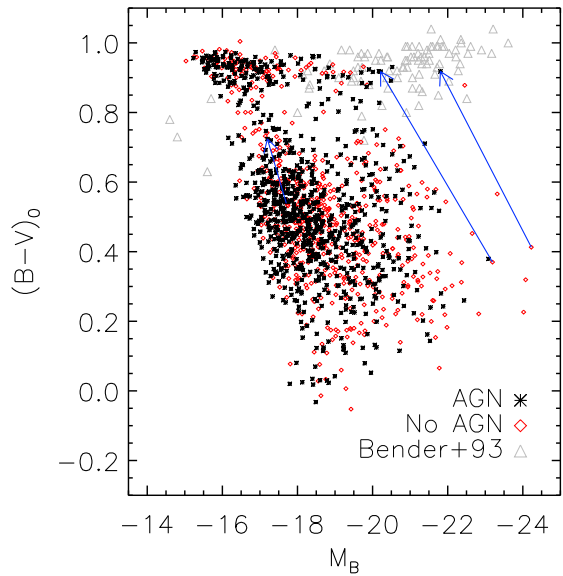}
\caption{Colour--magnitude relation for our simulated galaxies with (black stars) and without (red diamonds) AGN feedback.
Colours are measured within the central $1h^{-1}$ kpc.
Arrows show the motion of galaxies A, B, and C through the colour--magnitude plane when AGN feedback is included.
Observational data (grey triangles) are taken from \citet{bender93}.}
\label{fig:bbv}
\end{figure}

\begin{figure}
\centering
\includegraphics[width=0.48\textwidth,keepaspectratio]{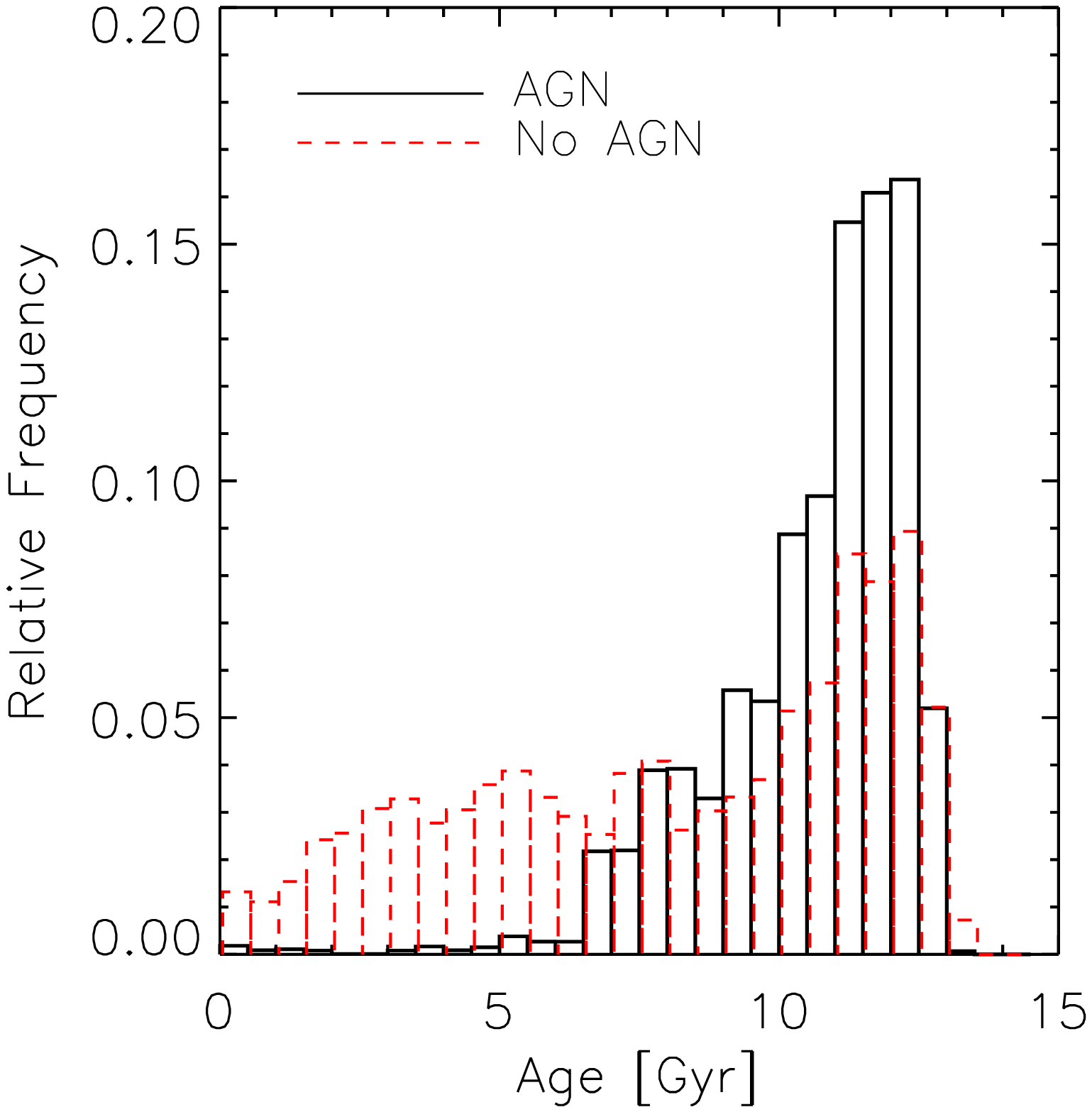}
\caption{Histograms of stellar age in the largest galaxy (A) in the simulation with (solid black) and without (dashed red) AGN feedback.}
\label{fig:histage}
\end{figure}

\begin{figure}
\centering
\includegraphics[width=0.48\textwidth,keepaspectratio]{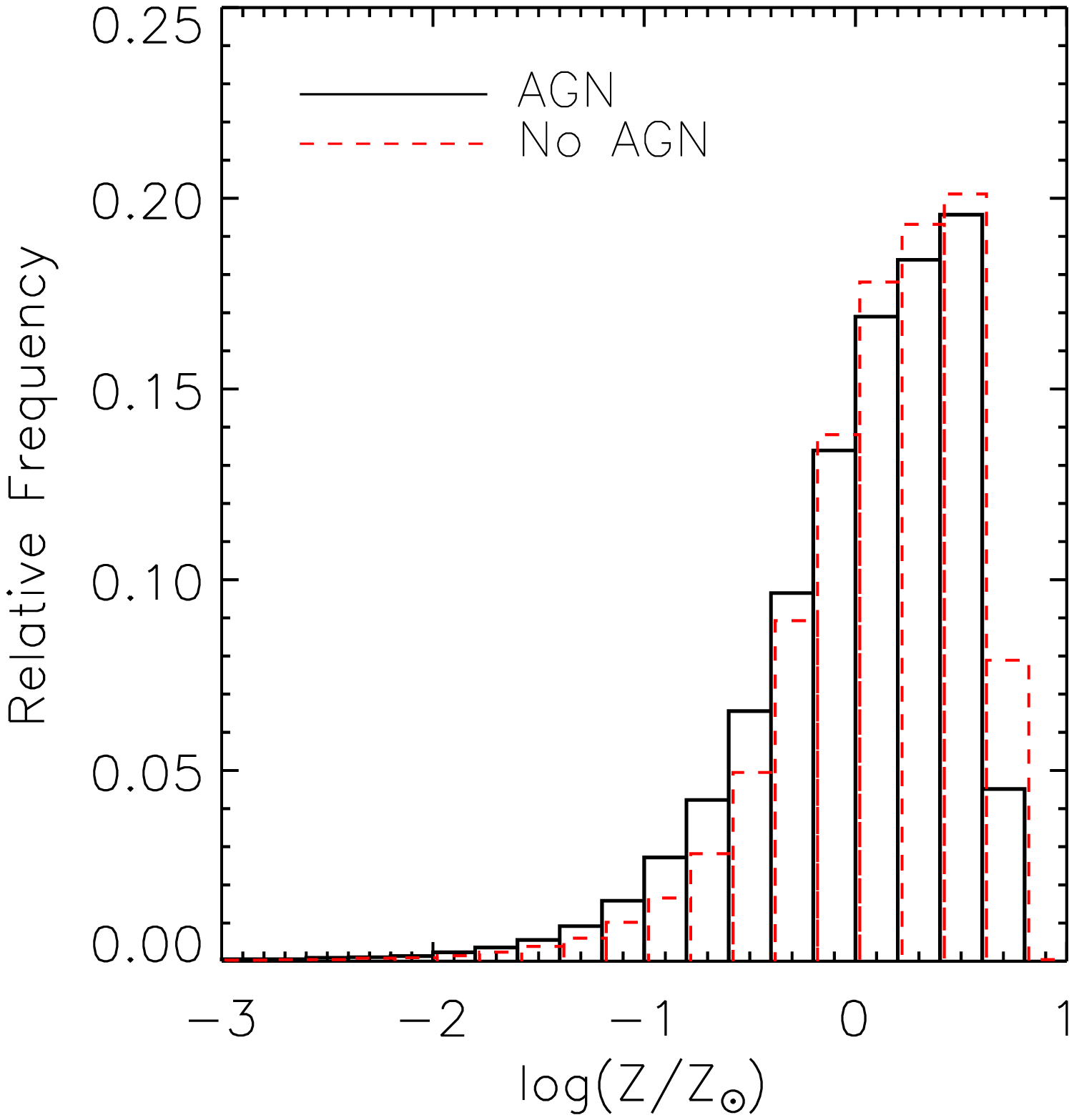}
\caption{Histograms of stellar metallicity in the largest galaxy (A) in the simulation with (solid black) and without (dashed red) AGN feedback.}
\label{fig:histmetal}
\end{figure}

\begin{figure}
\centering
\includegraphics[width=0.48\textwidth,keepaspectratio]{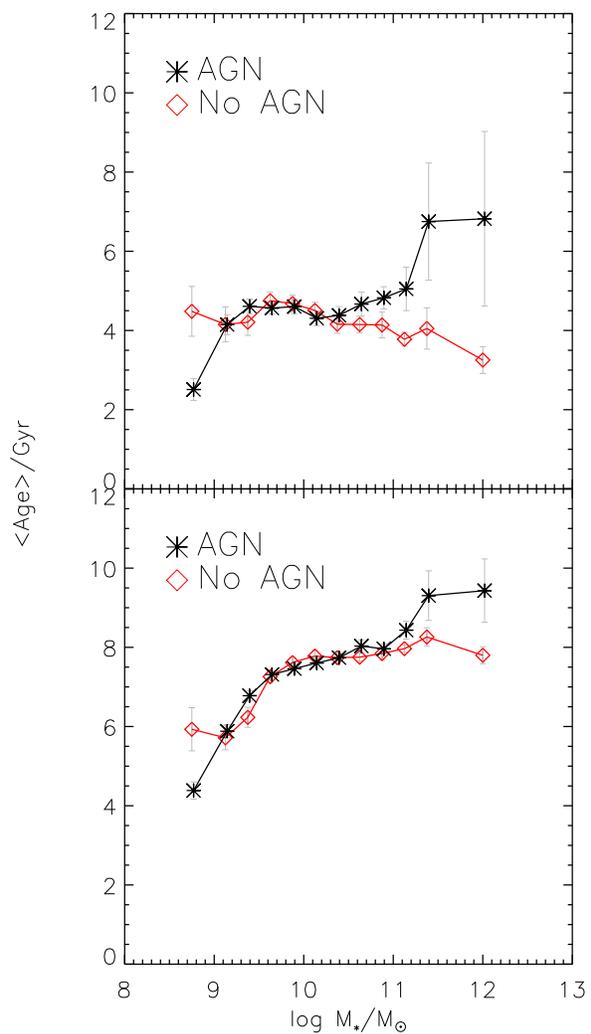}
\caption{Mean $V$ band luminosity-weighted (top panel) and mass-weighted (bottom panel) ages of galaxies as a function of stellar mass.
Error bars the standard error on the mean.
Black stars denote the simulation with AGN feedback, and the red diamonds without.}
\label{fig:galage}
\end{figure}

The colours of simulated galaxies are calculated by applying stellar population synthesis spectra \citep{bc03} to star particles as a function of age and metallicity.
Although the dust content can, in principal, be estimated for simulated galaxies \citep[e.g.,][]{delucia04}, the reddening due to dust is not included in this paper.
For this reason, we use optical bands where the bimodality \citep{kauffmann03} is less visible than for UV bands.

Fig. \ref{fig:bbv} shows central $B-V$ colour against $B$ band magnitude for our simulations; black stars with, and red diamonds without, AGN feedback.
Colour is measured within the central $1h^{-1}$ kpc for galaxies, for comparison with the observational data of \citet{bender93}.
Note that integrated colours across the whole galaxy tend to be bluer due to their colour gradient.
In both cases, a distinct red sequence ($(B-V)_0 \gtrsim0.8$), blue cloud ($(B-V)_0 \lesssim0.8$), and green valley ($(B-V)_0\sim0.8$) are seen.
The colours of massive galaxies become redder with AGN feedback, and consistent with the observations of \citet[][grey triangles]{bender93}.
At $M_{\rm B}>-18$, the simulated galaxies are on the extrapolated observed trend, but the number of these galaxies seems to be larger than observed, which may be affected by the resolution of our simulations.
This result is very similar to \citet{ck05}'s simulations of elliptical galaxies, although AGN feedback is not included there.

The movement of galaxies A, B, and C through the colour-magnitude plane is shown by the arrows.
Galaxy A becomes much redder and less luminous, with AGN.
Due to the significant reduction in star formation at low redshift (as shown in the top panel of Fig. \ref{fig:sfrabc}), and the dominance of very old stars ($>$10 Gyr) in the age distribution (Fig. \ref{fig:histage}), this galaxy moves onto the red sequence when AGN is included.
Most of the young stars form in the outer regions of the galaxy as a result of mergers with gas-rich satellites.
Fig. \ref{fig:histmetal} shows the metallicity distribution of this galaxy, which remains mostly unchanged with the inclusion of AGN, except for the most metal-rich bin. 	
The mass-weighted metallicity is not changed, but the luminosity-weighted metallicity is slightly decreased with AGN, and does not help in making colours redder.
If star formation is enhanced by AGN \citep[e.g.,][]{silk13}, the metallicity, and thus colour, could be larger.

Galaxy B also sees a significant reduction in star formation at late times when AGN feedback is included, to the extent that, by $z=0$, star formation has been shut off almost entirely (Fig. \ref{fig:sfrabc}).
Unlike in A, there are no significant mergers to provide gas, and so star formation almost totally stops once the BH has heated or removed the gas from the galaxy.
This allows B to move onto the red sequence.
In galaxy C, the BH grows to only $2\times10^5h^{-1}M_\odot$, and is not able to significantly affect star formation or colour.

Fig. \ref{fig:galage} provides another representation of the age difference.
More massive galaxies tend to be older, and in particular those with $\log M_*/M_\odot \gtrsim 11$ are older when AGN feedback is included.
This difference at $\log M_*/M_\odot \gtrsim 11$ is more pronounced for luminosity-weighted ages (top panel).
This AGN effect is in keeping with the findings of \citet{croton06}, but their galaxies were systematically older than our mass-weighted ages (bottom panel), which may be due to their cosmic SFR peaking at $z\sim4$.

\subsection{Mass--Metallicity Relation}
\label{sec:massmet}

\begin{figure}
\centering
\includegraphics[width=0.48\textwidth,keepaspectratio]{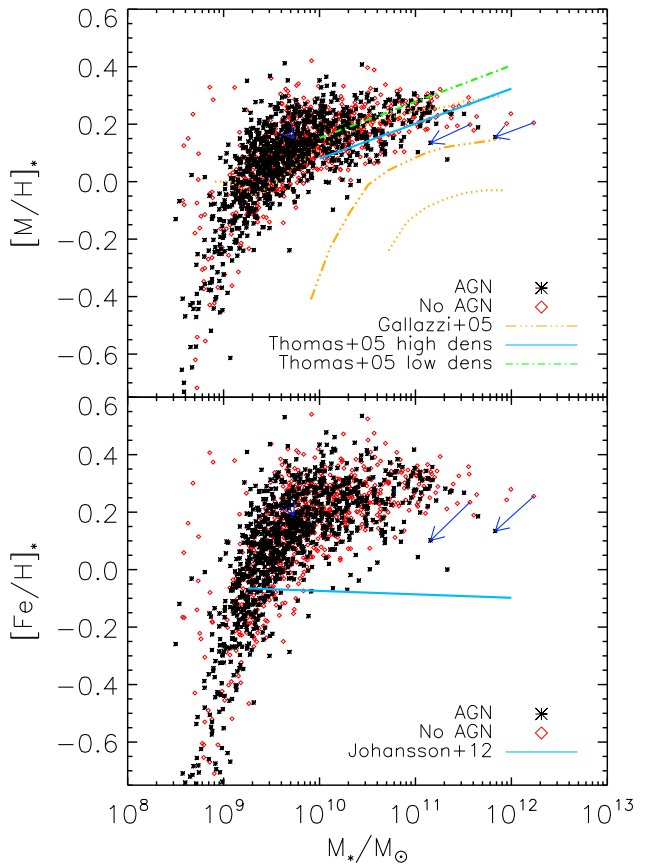}
\caption{Stellar mass metallicity relation for simulations with (black stars) and without (red diamonds) AGN feedback.
Top panel: $V$ band luminosity weighted total metallicity in $15h^{-1}$ kpc.
Observational data are taken from \citet[][solid blue and dashed green lines]{thomas05}, and \citet[][dot-dashed yellow line; scatter indicated by dotted lines]{gallazzi05}.
Bottom panel: $V$ band luminosity-weighted iron abundances in $15h^{-1}$ kpc.
The observed relation of \citet{johansson12} is also shown by the solid line.
Arrows show the movement of galaxies A, B, and C through the MZR when AGN feedback is included.}
\label{fig:massmetstar}
\end{figure}

\begin{figure}
\centering
\includegraphics[width=0.48\textwidth,keepaspectratio]{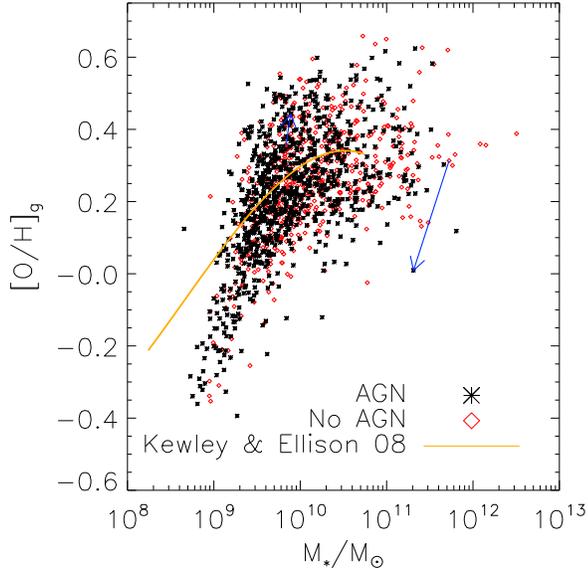}
\caption{Gas phase mass--metallicity relation.
SFR-weighted gas metallicity for the simulations with (black stars) and without (red diamonds) AGN feedback.
The solid line shows the SDSS observations, taken from \citet{kewley08} with \citet{kewley02}'s method, and arrows show how galaxies B and C move through the MZR when AGN feedback is included (there is no star-forming gas in A at $z=0$ when AGN feedback is included).
All metallicities are measured within $15h^{-1}$ kpc of the galactic centre.}
\label{fig:massmet}
\end{figure}

We show in the previous section that the stellar metallicity distribution in the most massive galaxy was not much affected by the inclusion of AGN (Fig. \ref{fig:histmetal}).
We now show, in the top panel of Fig. \ref{fig:massmetstar}, the stellar mass--metallicity relation (MZR) for galaxies in the simulations with and without AGN.
In order to compare with observations of optical absorption lines, we show the $V$ band luminosity-weighted stellar metallicity measured within the central $15h^{-1}$kpc of each galaxy.
We note that, due to the metallicity gradient within galaxies, slightly higher values of [M/H]$_*$ are obtained with a smaller radius, but this effect is much smaller than the scatter in the simulated relation.
There is no significant change in the MZR between the two simulations.
This is because AGN feedback quenches star formation most efficiently at low redshift, after the peak of star formation and chemical enrichment.
The simulated galaxies are in excellent agreement with the observations of \citet{thomas05}, with [M/H]$_*\sim0.2$ at $M_*\sim10^{11}M_\odot$.
The scatter of the simulations is also comparable to that of observations.
Note that \citet{gallazzi05}'s MZR of SDSS galaxies gives systematically lower metallicities at all masses, and even lower metallicities below $M_*\sim10^{10}M_\odot$.
Although these observations are subject to large scatter, most of our simulated galaxies lie above this MZR.

In the bottom panel of Fig. \ref{fig:massmetstar}, we show the stellar mass--iron abundance relation.
A stellar mass--iron abundance relation is seen in our simulations because, although there is less SN Ia enrichment in massive galaxies (as we show in the [$\alpha$/Fe] ratios of the following section), half of iron is produced from core-collapse supernovae, and thus the iron abundance also increases for massive galaxies.
We see that this relation mirrors closely the MZR in the upper panel, albeit with greater scatter.
However, there is no [Fe/H]$_*$--$M_*$ relation in \citet{johansson12}'s analysis of SDSS data, which may be due to the uncertainties in measuring elemental abundances from integrated spectra.

The slope of the MZR of our simulations is almost the same as in \citet{ck07}, but there is a systematic offset.
This is because the Kroupa IMF is used in this paper, rather than the Salpeter IMF in \citet{ck07}.
Although the slope of massive stars is very similar between these IMFs, the difference for low-mass stars changes the normalization, which changes the net yields from $0.6Z_\odot$ to $1Z_\odot$ \citep{ck11b}.
The slope of the MZR is generated by the mass-dependent galactic winds \citep{ck07}, which is also the case in this paper.

Arrows denote the movement of galaxies A, B, and C through the MZR.
Galaxies A and B, in which star formation is efficiently quenched, move down the observed relation to lower stellar mass and slightly lower ($\sim0.1$dex) metallicity.
Galaxy C, on which the effect of AGN feedback is minimal, moves only very slightly in either stellar mass or metallicity.

We show also, in Fig. \ref{fig:massmet}, the gas-phase MZR, using SFR-weighted oxygen abundances for comparison with observations that are estimated from emission lines.
We again compare the simulations with and without AGN, and include all of the gas in the central $15h^{-1}$kpc of each galaxy.
As with the stellar MZR, there is no significant difference between these simulations.
The simulated galaxies are consistent with SDSS observations, taken from \citet{kewley08} with \citet{kewley02}'s method (shown as the solid line), but there is much greater scatter than observed.
The scatter is also greater than has been found in other simulations \citep[e.g.][]{dave11}, though the adopted models differ both in their nucleosynthesis yields and the fact that our model includes feedback from AGN.

The movement of B and C through the gas-phase MZR is within the scatter, and is more complicated.
As well as influencing the enrichment history of galaxies, AGN feedback also drives galactic winds from the most massive galaxies.
This alters the metallicity of both the ISM and IGM compared to the case without AGN (Fig. \ref{fig:simevol}).
Consequently, there is no trend on the movement of our example galaxies (note that there is no star-forming gas in galaxy A at $z=0$, and so it has no arrow in Fig. \ref{fig:massmet}).

\subsection{[$\alpha$/Fe] - Velocity Dispersion Relation}
\label{sec:afe}

\begin{figure}
\centering
\includegraphics[width=0.48\textwidth,keepaspectratio]{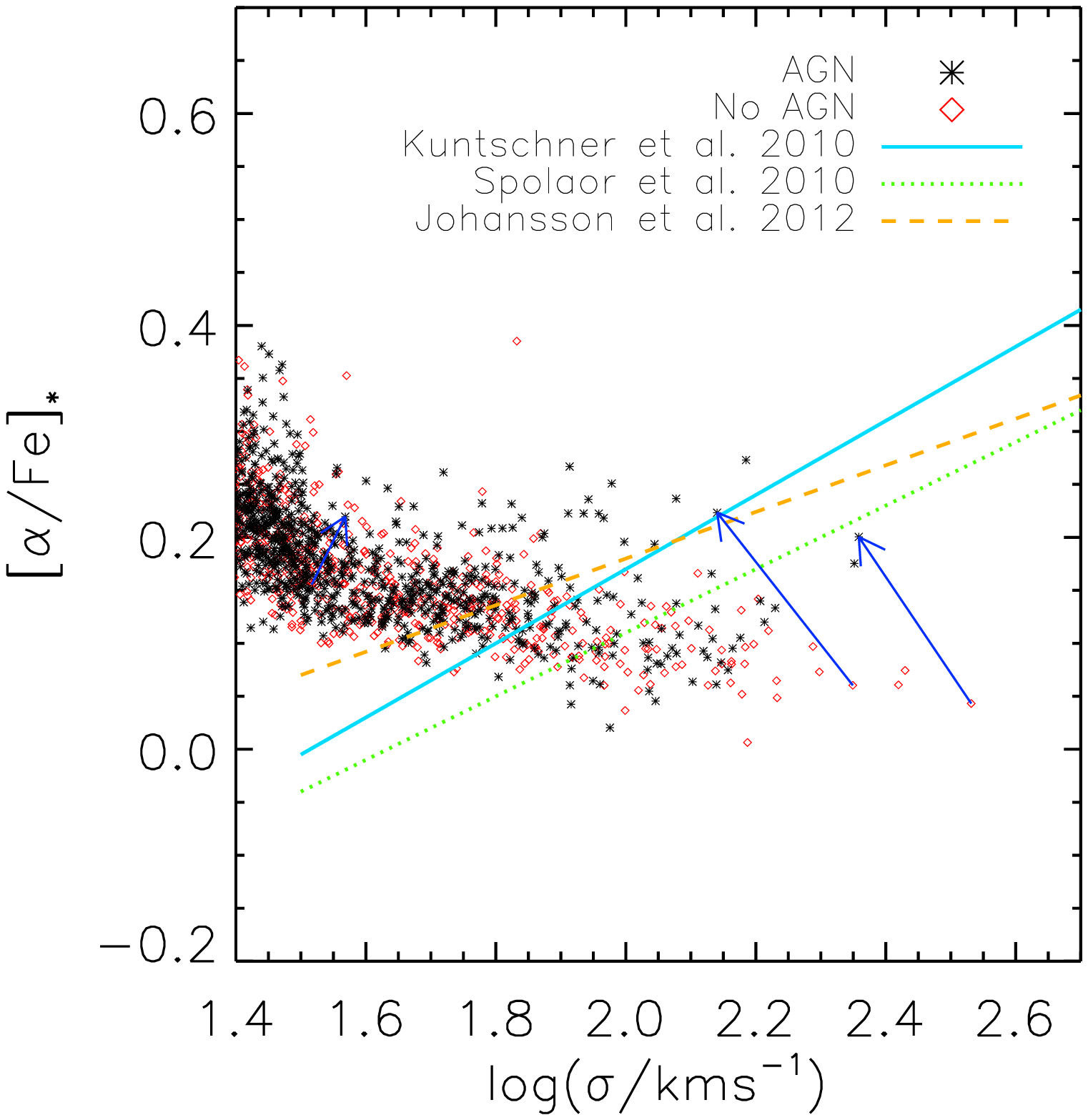}
\caption{[$\alpha$/Fe]--$\sigma$ relation for our simulations with (black stars) and without (red diamonds) AGN feedback.
Arrows denote the movement of galaxies A, B, and C through the [$\alpha$/Fe]--$\sigma$ relation when AGN feedback is included.
Observational data from \citet[][solid blue line]{kuntschner10}, \citet[][dotted green line]{spolaor10}, and \citet[][orange dashed line]{johansson12} are also shown.
$V$-band luminosity weighted [O/Fe] is measured within 1 effective radius for the simulated galaxies.}
\label{fig:afems}
\end{figure}

\begin{figure}
\centering
\includegraphics[width=0.48\textwidth,keepaspectratio]{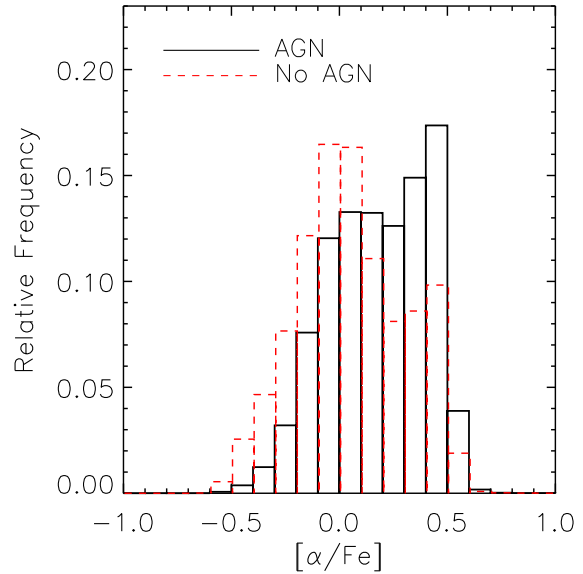}
\caption{Histograms of [$\alpha$/Fe] in the largest galaxy (A) in the simulation with (solid black line) and without (dashed red line) AGN feedback.}
\label{fig:histafe}
\end{figure}

The ratio of $\alpha$-elements to iron abundances gives information about the star formation timescale of a galaxy \citep[e.g.,][]{ck09}.
$\alpha$-elements are mostly produced by core-collapse supernovae with a short timescale ($\sim$Myr), while iron-peak elements are produced more by SNe Ia on longer timescales ($\sim$Gyr).
Therefore, at the early stages of galaxy formation, only core-collapse supernovae contribute to the chemical enrichment of the galaxy, leading to a higher $\alpha$/Fe ratio relative to the solar abundance.
At later stages, due to the delayed iron production by SNe Ia, [$\alpha$/Fe] decreases as a function of time.
The timescale of SNe Ia is an important issue, and depends on the progenitor model of SNe Ia; we take the best theoretical model that reproduces the observed [$\alpha$/Fe]--[Fe/H] relation in the Milky Way Galaxy (Section \ref{sec:modelstars}).

Fig. \ref{fig:afems} shows the [$\alpha$/Fe]--$\sigma$ relation for galaxies in our simulations with and without AGN feedback.
[$\alpha$/Fe] is measured, $V$ band luminosity-weighted, within the central effective radius of a galaxy.
Velocity dispersion, $\sigma$, is also measured within $1 R_{\rm e}$ of the centre.	
Without AGN, star formation takes place longer in more massive galaxies because of their deep potential well, and thus [$\alpha$/Fe] is lower for more massive galaxies, which is totally opposite to the observed relations.
This was also the case in \citet{ck07}, where AGN was not included.
With AGN, it is possible to quench the recent star formation in massive galaxies.
Although the effects of AGN feedback depend on environment, on average AGN feedback is more effective at quenching star formation in more massive galaxies.
These galaxies are made up of older stellar populations than their counterparts without AGN, as can be seen in Fig. \ref{fig:galage} (and is shown explicitly for the most massive galaxy in Fig. \ref{fig:histage}).
Since star formation is most suppressed at low redshift, the majority of stars formed from gas that had not been as enriched with iron as in the case without AGN.
As a result, more massive galaxies tend to have higher values of [$\alpha$/Fe] when BHs are included; this is shown explicitly for galaxy A in Fig. \ref{fig:histafe}.
The movement of galaxies A and B, in which AGN feedback is strong, in the [$\alpha$/Fe]--$\sigma$ plane also illustrate this, while C moves very little, since its star formation is not affected much.

Observational data show a tight relation between [$\alpha$/Fe] and galaxy mass for both central values \citep{thomas03,johansson12} and values within $R_{\rm e}$ \citep{kuntschner10,spolaor10}, although the slope and absolute values depend on the analysis of the observed absorption lines.
The aperture effect should be negligible as the observed radial gradient of [$\alpha$/Fe] is very small.

When AGN feedback is included, a number of galaxies move towards the observed relations for early-type galaxies \citep[][note that all simulated galaxies are displayed]{kuntschner10,spolaor10,johansson12}.
For $\log\sigma\gtsim2.1$, the relation is reversed when AGN is included, more in keeping with observations.
However, the scatter of the simulations is still larger than observed.
An improved model of AGN feedback may be required, or other possibilities have been discussed in \citet{ck05}, including (i) IMF, (ii) nucleosynthesis yields, (iii) binary fraction, and (iv) selective mass loss.

The [$\alpha$/Fe]--$\sigma$ relation was also shown with semi-analytic models with AGN feedback in \citet{arrigoni10} and \citet{gargiulo14}, both with a top-heavy IMF, and in \citet{calura09} with a variable IMF.
\citet{yates13} also requires modification of the SN Ia model.
However, our results suggest that the [$\alpha$/Fe]--$\sigma$ relation can basically be reproduced with our AGN feedback and standard stellar physics (IMF, binary fraction, and SN Ia model).

\subsection{Specific Star Formation Rate}

\begin{figure}
\centering
\includegraphics[width=0.48\textwidth,keepaspectratio]{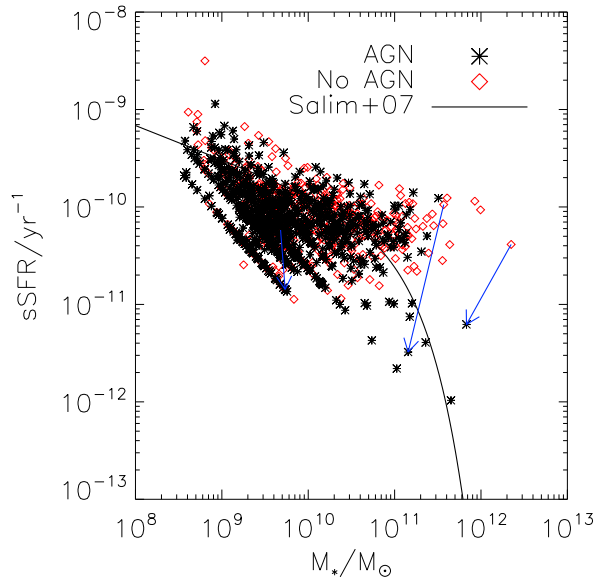}
\caption{Specific star formation rate against stellar mass for the simulations with (black stars) and without (red diamonds) AGN feedback.
Arrows show the effect of AGN feedback on galaxies A, B, and C.
The fit to observational data by \citet{salim07} is shown by the solid black line.}
\label{fig:ssfr}
\end{figure}

Finally, in Fig. \ref{fig:ssfr} we show the relationship between specific SFR (sSFR) and stellar mass for our simulated galaxies.
In both simulations, we see a trend for more massive galaxies to have lower sSFR, which is caused by supernova feedback.
This is in qualitative agreement with observations \citep[e.g.][]{salim07,ciambur13}.
With AGN feedback included, the most massive galaxies, such as A and B, show significant reduction in sSFR.
Smaller galaxies are less affected - for example, C moves very little in the sSFR\textendash$M_*$ plane.

Both observations and semi-analytic modeling \citep{ciambur13} find a bimodal distribution of sSFR at $z=0$, corresponding to low-mass galaxies with ongoing star formation, and quenched massive galaxies.
Our results do not show such a distinction, which may be because there are only a small number of galaxies with $\log M_*/M_\odot>11$ in our simulation box, and because the sSFR of $\ltsim 10^{-10}{\rm yr}^{-1}$ in less-massive galaxies ($\sim10^{10}M_\odot$) are affected by the resolution of our simulations (linear features in Fig. \ref{fig:ssfr}).


\section{Conclusions}
\label{sec:conc}

We have shown the effects of feedback from AGN on galaxy evolution using two cosmological hydrodynamical simulations, whose only difference is whether AGN feedback is included.
In our AGN model, black holes originate from Population {\sc iii} stars, in contrast to the merging origin of previous works, and grow via mergers and gas accretion.
Cosmic SFRs are greatly reduced at both high redshift ($z\gtrsim3$) and low redshift ($z\ltsim1$; Fig. \ref{fig:cosmicsfr}) with AGN feedback.
Since the black hole mass--galaxy mass relation is reproduced, massive galaxies host massive BHs.
These produce AGN-driven winds and quench the star formation there, causing them to be less massive, less compact, and redder by $z=0$ than their counterparts without AGN.
On the other hand, smaller galaxies do not host a supermassive BH and their star formation history is affected very little, but can get external enrichment from nearby AGN, depending on environment.
Nonetheless, galaxy mass--metallicity relations are not affected much since most stars form before AGN feedback becomes dominant.
The [$\alpha$/Fe] ratio of stellar populations is higher for massive galaxies due to their shorter star formation timescale, without modifying stellar physics (i.e., IMF, nucleosynthesis yields, SN Ia progenitors, and binary fraction).
Feedback from AGN plays an essential role in galaxy formation, namely in reproducing downsizing phenomena such as the colour-magnitude relation (Fig. \ref{fig:bbv}), specific star formation rates (Fig. \ref{fig:ssfr}), and the $\alpha$ enhancement in early type galaxies (Fig. \ref{fig:afems}).

Although our simulated galaxies are in better agreement with observations, there are still some mismatches seen, including: the luminosity function overpredicts the number of the most luminous galaxies; and the [$\alpha$/Fe]--$\sigma$ relation shows a larger scatter than observed. 
These suggest stronger feedback to suppress star formation at low redshift ($z\ltsim0.5$; Fig. \ref{fig:cosmicsfr}).
This cannot be achieved simply by changing the parameters of BH physics in our model, as these were chosen in \citet{pt14} to provide the best reproduction of observational constraints: the cosmic SFR; $M_{\rm BH}$--$\sigma$ relation; and galaxy size--mass relation.
There may be room for improvement by simultaneously changing the number of feedback neighbour particles, as discussed in detail in \citet{pt14}.
Alternatively, including smaller scale physics of AGN, kinetic feedback from jets, and induced star formation may be required, which will be investigated in future works.

\section*{Acknowledgements}
PT acknowledges funding from an STFC studentship, and thanks M. Hardcastle, D. Smith, J. Geach, and K. Coppin for useful discussions.
This work has made use of the University of Hertfordshire Science and Technology Research Institute high-performance computing facility.
This research made use of the DiRAC HPC cluster at Durham. DiRAC is the UK HPC facility for particle physics, astrophysics, and cosmology, and is supported by STFC and BIS.
CK acknowledges PRACE for awarding her access to resource ARCHER based in the UK at Edinburgh.
Finally, we thank V. Springel for providing GADGET-3.


\bibliographystyle{mn2e}
\bibliography{./refs}


\appendix

\bsp

\label{lastpage}

\end{document}